\documentclass[12pt]{article}

\usepackage{comment}
\usepackage[utf8]{inputenc}
\usepackage[margin=1in]{geometry}
\usepackage{amsmath}
\usepackage{multirow}
\usepackage{amssymb}
\usepackage{indentfirst}
\usepackage{amsfonts}
\newcommand\numberthis{\addtocounter{equation}{1}\tag{\theequation}}
\usepackage{scalerel}
\usepackage{natbib}
\usepackage{subcaption}
\usepackage[font=small, labelfont=bf]{caption}
\usepackage[colorlinks=true, allcolors=blue]{hyperref}
\usepackage[dvipsnames,svgnames]{xcolor}
\usepackage[english]{babel}
\usepackage[mathscr]{euscript}

\usepackage{booktabs} 

\usepackage{fancyvrb} 

\usepackage{siunitx}

\usepackage{xpatch}
\xapptocmd\normalsize{%
 \abovedisplayskip=4pt plus 2pt minus 2pt
 \belowdisplayskip=4pt plus 2pt minus 2pt
}{}{}

\usepackage{thmtools}
\usepackage{mathtools}

\usepackage[capitalize, nameinlink, sort]{cleveref}
\crefname{thm}{Theorem}{Theorems}
\crefname{cor}{Corollary}{Corollary}
\crefname{lem}{Lemma}{Lemmas}
\crefname{asu}{Assumption}{Assumptions}
\crefname{rmk}{Remark}{Remarks}
\crefname{defn}{Definition}{Definitions}
\crefname{thmlisti}{Theorem}{Theorems}
\crefname{asulisti}{Assumption}{Assumptions}

\usepackage{enumitem}

\usepackage{letltxmacro}

\newlist{thmlist}{enumerate}{1}
\setlist[thmlist]{label=(\roman*), ref=\thethm\,(\roman*)}

\newlist{asulist}{enumerate}{1}
\setlist[asulist]{label=(\roman*), ref=\theasu\,(\roman*)}

\usepackage{thmtools}
\declaretheorem[
    name=Theorem,
    Refname={Theorem,Theorems}]{thm}

\declaretheorem[
    name=Assumption,
    Refname={Assumption,Assumptions}]{asu}
\declaretheorem[
    name=Remark,
    Refname={Remark,Remarks},
    numberwithin=section]{rmk}

\numberwithin{equation}{section}

\newcommand{\cee}{\text{CEE}}

\newcommand{\bA}{\bar{A}}
\newcommand{\ba}{\bar{a}}
\newcommand{\RR}{\mathbb{R}}
\newcommand{\PP}{\mathbb{P}}
\newcommand{\tp}{\tilde{p}}
\newcommand{\tU}{\tilde{U}}

\newcommand{\cT}{\mathcal{T}}
\newcommand{\bp}{{\boldsymbol{p}}}
\newcommand{\bpi}{{\boldsymbol{\pi}}}

\newcommand{\prob}{\text{Pr}}

\newcommand{\EE}{\mathbb{E}}

\usepackage{dsfont}

\usepackage[ruled]{algorithm2e}

\def\spacingset#1{\renewcommand{\baselinestretch}%
{#1}\small\normalsize} \spacingset{1}

\begin{document}

  \title{\bf Doubly Robust Estimation of Causal Excursion Effects in Micro-Randomized Trials with Missing Longitudinal Outcomes \vspace*{0.5cm}}
   \date{\vspace{-2ex}} 
  \author{Jiaxin Yu \\ 
   Department of Statistics, University of California, Irvine    \vspace*{0.3cm}\\ 
   Tianchen Qian \\
   Department of Statistics, University of California, Irvine}

\maketitle

\vspace{0.2cm}

\begin{abstract}
Micro-randomized trials (MRTs) are increasingly utilized for optimizing mobile health interventions, with the causal excursion effect (CEE) as a central quantity for evaluating interventions under policies that deviate from the experimental policy. However, MRT often contains missing data due to reasons such as missed self-reports or participants not wearing sensors, which can bias CEE estimation. In this paper, we propose a two-stage, doubly robust estimator for CEE in MRTs when longitudinal outcomes are missing at random, accommodating continuous, binary, and count outcomes. Our two-stage approach allows for both parametric and nonparametric modeling options for two nuisance parameters: the missingness model and the outcome regression. We demonstrate that our estimator is doubly robust, achieving consistency and asymptotic normality if either the missingness or the outcome regression model is correctly specified. Simulation studies further validate the estimator’s desirable finite-sample performance. We apply the method to HeartSteps, an MRT for developing mobile health interventions that promote physical activity.
\end{abstract}

\noindent%
\vfill

\newpage

\spacingset{1.9} 

\section{Introduction}
\label{sec:introduction}

Mobile health (mHealth) interventions, such as push notifications delivered through smartphones and wearables, are being developed in various domains. Micro-randomized trial (MRT) is an experimental design for developing and optimizing these interventions \citep{klasnja2015microrandomized, dempsey2015randomised}. In an MRT, each individual is repeatedly randomized among treatment options hundreds or thousands of times, resulting in longitudinal data with time-varying treatments. A key quantity of interest in MRT analysis is the causal excursion effect (CEE). The CEE measures whether an intervention is effective and how its effect changes over time or interacts with the individual’s context, which are crucial for optimizing interventions. Semiparametric methods have been proposed for estimating the CEE when the longitudinal outcomes are always observed \citep{boruvka2018assessing,qian2021estimating,cheng2023efficient}.

Missing data remains a critical and nearly universal challenge in MRTs, with few principled solutions currently available. Common causes of missingness include missed self-reports and participants not wearing sensors \citep{seewald2019practical}. Current practices for handling missing data in MRTs, such as imputation with zero values \citep{klasnja2019efficacy} or modal imputation \citep{bell2023notifications}, are not principled and can lead to biased estimates. \citet{shi2023meta} proposed an inverse probability weighted estimator, but their approach requires correctly specified missingness model and may suffer from large variance \citep{lee2011weight}.

We propose an estimator for the CEE in MRTs where longitudinal outcomes are missing at random. We focus on CEE defined with identify or log link, which covers common longitudinal outcome types including continuous, binary, and count. Our estimator is doubly robust in the sense that it is consistent and asymptotically normal if one of the two nuisance models, a propensity score model for the missingness and an outcome regression model, is correctly specified. Furthermore, the proposed estimator allows the nuisance models to be fitted by data-adaptive or machine learning methods which typically converge at a slower-than-parametric rate. We established the asymptotic properties and validated them via simulations. The method is applied to the HeartSteps MRT \citep{klasnja2015microrandomized}, which aims to promote physical activity of participants.

Being the first doubly robust estimator for CEE with missing outcomes, our method builds upon the extensive literature on semiparametric and doubly robust estimation with missing data \citep{robins1994estimation, scharfstein1999adjusting,bang2005doubly,tsiatis2006semiparametric,vermeulen2015bias}. We also incorporate recent advancements in data-adaptive methods for causal inference \citep{chernozhukov2018double,kennedy2022semiparametric}.

This paper is organized as follows. In Section \ref{sec: Preliminaries}, we define notations and CEE. In Section \ref{sec: Doubly Robust CEE}, we first review methods for estimating CEE in the absence of missing data, and then present the proposed method for settings with missing outcomes. In Section \ref{sec: Asymptotic Theory}, we establish the asymptotic properties and double robustness of the proposed estimator. Simulation studies are presented in Section \ref{sec: Simulation}. The method is illustrated using the HeartSteps study in Section \ref{sec: Application}. We conclude with a discussion in Section \ref{sec: Discussion}.

\section{Preliminaries}
\label{sec: Preliminaries}

\subsection{Notation}

Consider an MRT that includes observations from $n$ individuals, each enrolled for $T$ decision points at which treatments are randomized. Variables without subscript $i$ represent observations from a generic individual. Let $X_t$ be the observation from the individual between decision points $t-1$ and $t$. Let $A_t$ represent the treatment assignment at decision point $t$, with 1 indicating treatment and 0 indicating no treatment. Let $I_t \in X_t$ denote the availability indicator, where $A_t = 0$ deterministically if $I_t = 0$ (which can occur, for instance, if the individual is driving for safety reasons). Note that a decision point being unavailable is distinct from variables being missing. The overbar is used to denote a sequence of variables up to a decision point; for example, $\bA_t = (A_1,..., A_t)$.  Information accrued up to time $t$ is represented by $H_t = (X_1,A_1,X_2,A_2,...,X_{t-1}, A_{t-1}, X_t) = (\bar{X}_t, \bar{A}_{t-1})$. The randomization probability for $A_t$ can depend on $H_t$ and is denoted by $p_t(H_t) := \prob(A_t = 1|H_t)$, and we sometimes omit $H_t$ to simply write $p_t$. 

The proximal outcome following the treatment assignment at time $t$, denoted by $Y_{t, \Delta}$, is a known function of the individual’s data in a subsequent time window of length $\Delta$, where $\Delta \geq 1$ is a fixed positive integer; that is, $Y_{t,\Delta} = y(X_{t+1}, A_{t+1},...,X_{t+\Delta-1}, A_{t+\Delta-1}, X_{t+\Delta})$ for some known function $y(\cdot)$. A researcher would set $\Delta = 1$ if they are interested in how the treatment impacts the most immediate outcome, which is the most common practice \citep{klasnja2019efficacy,bell2023notifications}. A $\Delta$ greater than 1 allows one to assess delayed effects. We define these effects rigorously in Section \ref{subsec: potential outcomes and causal effect}.

We consider the setting where $X_t$ and $A_t$ are always observed and $Y_{t, \Delta}$ can be missing. Let $R_{t, \Delta}$ denote whether $Y_{t, \Delta}$ is observed (in which case $R_{t, \Delta} = 1$) or missing ($R_{t, \Delta} = 0$). We assume that the data from different individuals are independent and identically distributed samples from an unknown distribution. Throughout, we represent random variables or vectors with uppercase letters, and their realized values with lowercase letters.

We use $\EE$ to denote expectation and $\PP_n$ to denote the empirical average over all individuals. For a positive integer $n$, We use $[n]$ to denote the set $\{1,2,...,n\}$. We use superscript $^\star$ to denote quantities corresponding to the true data generating distribution. We use $\|\cdot\|$ to denote the $L_2$ norm, i.e. $\|f(O)\| = \{\int f(o)^2 d P(o)\}^{1/2}$ for any function $f$ of the observed data $O$. For a vector $\alpha$ and a vector-valued function $f(\alpha)$, $\partial_\alpha f(\alpha) := \partial f(\alpha)/\partial \alpha^T$ denotes the matrix where the $(i,j)$-th entry is the partial derivative of the $i$-th entry of $f$ with respect to the $j$-th entry of $\alpha$.

\subsection{Potential Outcomes and Causal Excursion Effect}
\label{subsec: potential outcomes and causal effect}

To define causal effects, we use the potential outcomes framework \citep{rubin1974estimating, robins1986new}. For an individual, let $X_t(\ba_{t-1})$ be the observation that would have been observed at decision point $t$ if the individual had been assigned a treatment sequence of $\ba_{t-1}$. The potential outcome of $H_t$ under $\ba_{t-1}$ is $H_t(\ba_{t-1}) = \{X_1, A_1, X_2(a_1),A_2(a_1),...,A_{t}(\ba_{t-1}), X_{t+1}(\ba_{t})\}$. The potential outcome for the proximal outcome is $Y_{t,\Delta}(\bar{a}_{t+\Delta-1})$.

A causal excursion effect (CEE) is a contrast between the potential outcomes under two ``excursions'', i.e., treatment regimes that deviate from the treatment regime in the MRT \citep{boruvka2018assessing,qian2021estimating}. Here treatment regimes refer to how the treatment is sequentially assigned at each decision point depending on the history information \citep{murphy2003optimal}. Formally, let $\bp$ denote the treatment regime in the MRT where for $t \in [T]$ the randomization probability of $A_t$ is $p_t(H_t)$, and let $\bpi$ denote some other regime where for $t \in [T]$ the randomization probability of $A_t$ is $\pi_t(H_t)$. We require $\bpi$ to be compatible with the availability, i.e., $\pi_t(H_t) = 0$ if $I_t = 0$. Let $S_t \subset H_t$ denote the effect modifiers of interest, chosen by the researcher depending on the scientific question. Given a link function $g(\cdot)$ that is either the identity or log function, the CEE of $A_t$ on $Y_{t, \Delta}$ moderated by $S_t$ is
\begin{align*}
    \cee_{\bp, \bpi; \Delta}(t; s) &= g[\EE_{\bA_{t-1} \sim \bp, \bA_{t+1:t+\Delta-1} \sim \bpi}\{Y_{t, \Delta}(\bA_{t-1}, 1, \bA_{t+1:t+\Delta-1}) \mid S_t(\bA_{t-1}) = s, I_t(\bA_{t-1}) = 1 \} ] \\
    & ~~~ - g[\EE_{\bA_{t-1} \sim \bp, \bA_{t+1:t+\Delta-1} \sim \bpi}\{Y_{t, \Delta}(\bA_{t-1}, 0, \bA_{t+1:t+\Delta-1} \mid S_t(\bA_{t-1}) = s, I_t(\bA_{t-1}) = 1 \} ].
    \numberthis
    \label{eq: causal excursion effect}
\end{align*}
The right hand side of \eqref{eq: causal excursion effect} is the contrast in potential outcomes under the two excursions: $(\bA_{t-1}, 1, \bA_{t+1:t+\Delta-1})$ and $(\bA_{t-1}, 0, \bA_{t+1:t+\Delta-1})$, where $\bA_{t-1}$ (the treatments prior to $A_t$) follow the MRT regime $\bp$, and $\bA_{t+1:t+\Delta-1} = (A_{t+1},...,A_{t+\Delta-1})$ (the treatments after $A_t$ but prior to the measurement of $Y_{t,\Delta}$) follow the treatment regime $\bpi$. The choice of $\bpi$ impacts the interpretation of the CEE; for example, setting $\pi_t = p_t$ for all $t$ yields the lagged effects considered in \citet{boruvka2018assessing}, and setting $\pi_t = 0$ for all $t$, which means that $\bA_{t+1:t+\Delta-1} = 0$, yields the effects considered in \citet{dempsey2020stratified} and \citet{qian2021estimating}. When $\Delta = 1$, i.e., when the interest is in the treatment effect on the most immediate outcome $Y_{t, 1}$, $\bpi$ becomes irrelevant in \eqref{eq: causal excursion effect} and the CEE becomes a contrast between $Y_{t, 1}(\bA_{t-1}, 1)$ and $Y_{t, 1}(\bA_{t-1}, 0)$. This is the most common scenario in applications \citep{klasnja2019efficacy,qian2022microrandomized,bell2023notifications}.

The CEE \eqref{eq: causal excursion effect} is moderated by the effect modifiers of interest $S_t$. A common choice of $S_t$ is the empty set, where the CEE is a fully marginal effect over all possible effect modifiers. Alternatively, one may include in $S_t$ past treatments or past covariates/outcomes to assess interactions between treatments at different decision points or effect modification by past covariates/outcomes. The CEE is conditional on $I_t(\bA_{t-1}) = 1$, the individual being available at the decision point. This is necessary for the CEE to be nonparametrically identified, and it also makes the CEE relevant in the mHealth context; after all, safe and ethical interventions should only be sent at times deemed appropriate \textit{a priori}, i.e., when the individual is available, and thus the relevant causal effect should only concern such time points. Conditioning on availability is connected to the notion of feasible dynamic treatment regimes \citep{robins2004optimal,wang2012evaluation}.

\section{A Doubly Robust Estimator under Missing at Random}
\label{sec: Doubly Robust CEE}

\subsection{Estimator with Fully Observed Outcomes}

We consider the estimation of an unknown parameter $\beta$ whose true value $\beta^\star \in \RR^p$ satisfies
\begin{align}
  \cee_{\bp, \bpi; \Delta}(t; s) = f_t(s)^T\beta^\star \text{ for } t \in [T], \label{eq:cee-model}
\end{align}
where $f_t(\cdot)$ is a pre-specified function. For example, $f_t(s)^T\beta$ can include a linear model with effect modifiers $s$ and basis functions of $t$. We make the following assumptions.
\begin{asu}
    \label{assu: consistency, Positivity, Sequential ignorability.}
    \begin{asulist}
    \item \label{assu: consistency} (Stable unit treatment value assumption [SUTVA]).
    The observed data is the same as the potential outcome under the observed treatment assignment, and one's potential outcomes are not affected by others' treatment assignments. Specifically, $R_{t, \Delta} = R_{t, \Delta}(\bA_{t + \Delta - 1}), Y_{t, \Delta} = Y_{t, \Delta}(\bA_{t + \Delta - 1})$, and $X_t = X_t(\bA_{t-1})$ for $t \in [T]$. 
    \item \label{assu: positivity} (Positivity).
    There exists $c > 0$ such that $c < \prob(A_t = 1|H_t, I_t = 1) < 1 - c$ almost surely for all $t\in[T]$. 
    \item \label{assu: Sequential ignorability} (Sequential ignorability).
    The potential outcomes $\{X_{t+1}(\ba_{t}),  A_{t+1}(\ba_{t}), \ldots,  X_{T+1}(\ba_{T})\}$ are independent of $A_t$ conditional on $H_t$ for $t \in [T]$. 
    \end{asulist}
\end{asu}  

In a micro-randomized trial, because the treatment is sequentially randomized with known probabilities bounded away from 0 and 1, Assumptions \ref{assu: consistency, Positivity, Sequential ignorability.} (ii) and (iii) are satisfied by trial design. Note that Assumption \ref{assu: consistency, Positivity, Sequential ignorability.} (iii) implies the sequential ignorability concerning $Y_{t,\Delta}$ and $R_{t,\Delta}$ because they are functions of $(X_{t+1}, A_{t+1},...,X_{t+\Delta-1}, A_{t+\Delta-1}, X_{t+\Delta})$. Assumption \ref{assu: consistency, Positivity, Sequential ignorability.} (i) may fail to hold if there is peer influence or social interaction between individuals \citep{hudgens2008toward,shi2023assessing}. To maintain the focus of this paper, we do not consider such settings here. In the Section A of Supplementary Material, we show that under Assumption \ref{assu: consistency, Positivity, Sequential ignorability.}, the CEE \eqref{eq: causal excursion effect} is identified as
\begin{align*}
    \cee_{\bp, \bpi; \Delta}(t; s) &= g[\EE_{\bA_{t-1}\sim\bp}\{\EE_{\bA_{t+1:t+\Delta-1}\sim\bp}( W_{t, \Delta} Y_{t, \Delta} \mid A_t = 1, H_t) \mid S_t, I_t = 1 \} ] \\
    & ~~~ - g[\EE_{\bA_{t-1}\sim\bp}\{\EE_{\bA_{t+1:t+\Delta-1}\sim\bp}( W_{t, \Delta} Y_{t, \Delta} \mid A_t = 0, H_t) \mid S_t, I_t = 1 \} ],
    \numberthis
    \label{eq: observed causal excursion effect}
\end{align*}
where $W_{t, \Delta} := \prod_{u = t+1}^{t + \Delta - 1} \big\{\frac{\pi_u(H_u)}{p_u(H_u)}\big\}^{A_u} \big\{\frac{1 - \pi_u(H_u)}{1 - p_u(H_u)}\big\}^{1 - A_u}$  can be interpreted as a change of probability from $p_u$ to $\pi_u$ for future assignment $\bA_{t+1:t+\Delta-1}$, and we set $W_{t, \Delta} := 1$ if $\Delta = 1$.

We describe an estimating function for $\beta^\star$ in \eqref{eq:cee-model} that is unbiased under Assumption \ref{assu: consistency, Positivity, Sequential ignorability.} when the outcome is always observed. For each $t \in [T]$, let $\mu_t(h, a)$ be a nuisance function with truth $\mu_{t}^\star(h, a) := \EE(Y_{t, \Delta} \mid H_t = h, A_t = a)$. We define the stabilized inverse probability weight $W_t := \big\{\frac{\tp_t(S_t)}{p_t(H_t)}\big\}^{A_t} \big\{\frac{1 - \tp_t(S_t)}{1 - p_t(H_t)}\big\}^{1 - A_t}$ for $t \in [T]$, where the numerator probability $\tp_t(S_t)$ is a nuisance function that takes value in $(0,1)$ and can be estimated, for example, using a logistic regression of $A_t$ on $S_t$. Note that $W_t$ is distinct from $W_{t,\Delta}$. To describe the estimator with fully observed outcomes, we use shorthand notation $\mu_{at} := \mu_t(H_t, a)$ for $a \in \{0,1\}$, $p_t := p_t(H_t)$, and $\tp_t := \tp_t(S_t)$ for $t \in [T]$. Let $U_t(\beta, \mu_t, \tp_t) := I_t W_t W_{t,\Delta} \epsilon_t (A_t - \tp_t) f_t(S_t)$, where
\begin{align*}
    \epsilon_t := \begin{cases}
        Y_{t, \Delta} - (A_t + p_t - 1 ) f_t(S_t)^T \beta - (1 - p_t) \mu_{1t} - p_t \mu_{0t} & \text{ if $g$ is identity}, \\ 
        e^{-A_t f_t(S_t)^T\beta}Y_{t, \Delta} - (1 - p_t) e^{-f_t(S_t)^T \beta}\mu_{1t} - p_t \mu_{0t} & \text{ if $g$ is log}.
    \end{cases}
\end{align*}
A two-stage estimator for $\beta$ can be constructed by first obtaining nuisance function estimators $\hat\mu_t$ and $\hat\tp_t$, and then solving for $\hat\beta$ from the estimating equation $\PP_n \sum_{t=1}^T U_t(\hat\beta, \hat\mu_t, \hat\tp_t) = 0$. When $g$ is identity, $\bpi = \bp$, and $\epsilon_t$ is replaced by $Y_{t,\Delta} - (A_t - \tp_t)f_t(S_t)^T\beta - \eta_t(H_t)^T\alpha$ for a pre-specified function $\eta_t(H_t)$ (often called control variables), $\hat\beta$ corresponds to a two-stage version of the weighted and centered least squares method in \citet{boruvka2018assessing}. When $g$ is log, $\pi_t = 0$ for all $t$, and $\epsilon_t$ is replaced by $e^{-A_t f_t(S_t)^T\beta}Y_{t, \Delta} - e^{\eta_t(H_t)^T\alpha}$ for pre-specified control variables $\eta_t(H_t)$, $\hat\beta$ correspond to a two-stage version of the estimator for marginal excursion effect in \citet{qian2021estimating}. \citet{cheng2023efficient} showed that the two-stage estimator $\hat\beta$ for fully observed outcomes is consistent and asymptotically normal for arbitrary choice of $\hat\mu_t$ and $\hat\tp_t$ when $\Delta = 1$, and the particular form of $\epsilon_t$ was chosen to improve the efficiency of $\hat\beta$ over \citet{boruvka2018assessing} and \citet{qian2021estimating}.


\subsection{Proposed Estimator For Outcomes Missing At Random}
\label{subsec: Proposed Estimator For Outcomes Missing At Random}

We make additional assumptions when the outcome can be missing.

\begin{asu} \label{assu: positivity of missingness probability stated in the main paper}
    \begin{asulist}
    \item  (Positivity for $R_{t,\Delta}$).
    There exists $c > 0$ such that $\prob(R_{t, \Delta} = 1|H_t, A_t) $ $> c$ almost surely for all $t\in[T]$. 
    \item (Missing at random given history information).
    $Y_{t, \Delta} \perp R_{t, \Delta} \mid H_t, A_t$ for all $t\in[T]$.
    \end{asulist}
\end{asu}

For each $t \in [T]$, let $e_t(H_t, A_t)$ be a nuisance function with truth $e_t^\star(H_t, A_t) := $ $\prob(R_{t, \Delta} $ $= 1 \mid H_t, A_t)$. We propose the estimating function $\sum_{t=1}^T \tU_t(\beta, e_t, \mu_t, \tp_t)$ with
\begin{align}
    \tU_t(\beta, e_t, \mu_t, \tp_t) := \frac{R_{t, \Delta}}{e_t(H_t, A_t)} \ U_t (\beta, \mu_t, \tp_t) - \frac{R_{t, \Delta} - e_t(H_t, A_t)}{e_t(H_t, A_t)} \ \EE\{U_t (\beta, \mu_t, \tp_t) \mid H_t, A_t\}. \label{eq:u-tilde-general}
\end{align}
The first term in \eqref{eq:u-tilde-general} is $U_t$ weighted by the inverse probability of $Y_{t,\Delta}$ being observed. The second term in \eqref{eq:u-tilde-general} is an augmentation term that yields the double robustness (Section \ref{sec: Asymptotic Theory}) based on the seminal work of \citet{robins1994estimation}. 
Algebraic calculation yields
\begin{align}
    \tU_t(\beta, e_t, \mu_t, \tp_t) = I_t W_t W_{t, \Delta} \Big[\frac{R_{t, \Delta}}{e_t(H_t, A_t)}\{Y_{t, \Delta} - A_t\mu_{1t} - (1-A_t)\mu_{0t} \} \nonumber\\
 + (A_t + p_t - 1)\{\mu_{1t} - \mu_{0t} - f_t(S_t)^T \beta\} \Big] (A_t - \tp_t)f_t(S_t) \label{eq:u-tilde-continuous}
\end{align}
if the link function $g$ is identity, and
\begin{align}
    \tU_t(\beta, e_t, \mu_t, \tp_t) 
    =  I_t W_t W_{t,\Delta} \Big[ \frac{R_{t, \Delta}}{e_t(H_t, A_t)} e^{-A_t f_t(S_t)^T\beta} \{Y_{t, \Delta} - A_t\mu_{1t} - (1-A_t)\mu_{0t}\} \nonumber\\
    + (A_t + p_t - 1)\{e^{-f_t(S_t)\beta} \mu_{1t} - \mu_{0t}\} \Big](A_t - \tp_t)f_t(S_t) \label{eq:u-tilde-binary}
\end{align}
if the link function $g$ is $\log$.

We propose a two-stage implementation for $\hat\beta$ by first estimating the nuisance parameters in $\tU_t(\beta, e_t, \mu_t, \tp_t)$ and then solve an estimating equation for $\hat\beta$, detailed in Algorithm \ref{algo:est}.

\begin{algorithm}[htbp]
    \caption{Causal excursion effect estimator $\hat\beta$ for outcomes missing at random}
    \label{algo:est}
    \spacingset{1.5}
    \vspace{0.3em}
    Given the causal excursion effect model \eqref{eq:cee-model}:
    
    \textbf{Stage 1:} Fit $\prob(R_{t, \Delta} = 1 \mid H_t, A_t)$, $\EE (Y_{t, \Delta} \mid H_t, A_t, I_t = 1)$, and $\prob (A_t = 1 \mid S_t, I_t = 1)$ for $t \in [T]$. Denote the fitted models by $\hat{e}_t(H_t, A_t)$, $\hat\mu_t(H_t, A_t)$, and $\hat\tp_t(S_t)$, respectively.

    \textbf{Stage 2:} Obtain $\hat\beta$ by solving $\PP_n \sum_{t=1}^T \tU_t(\beta, \hat{e}_t, \hat\mu_t, \hat\tp_t) = 0$ with $\tU_t$ defined in \eqref{eq:u-tilde-continuous} if $g$ is identity and \eqref{eq:u-tilde-binary} if $g$ is $\log$.
    \vspace{0.3em}
\end{algorithm}


\begin{rmk}
    A wide variety of flexible methods can be used in Stage 1 of Algorithm \ref{algo:est} to fit the nuisance parameters, including parametric methods and nonparametric methods such as kernel regression, spline methods with complexity penalties, and ensemble methods \citep{fan1996local, ruppert2003semiparametric, van2011cross}. Depending on the model fitting method in Stage 1, the required conditions for obtaining a consistent and asymptotically normal $\hat\beta$ in Stage 2 are slightly different. We explain this in detail in Section \ref{sec: Asymptotic Theory}.
\end{rmk}

\section{Asymptotic Theory}
\label{sec: Asymptotic Theory}

We make the following assumptions for establishing the asymptotic normality of the two-stage estimator in Algorithm \ref{algo:est}. Proof of the theorems is given in Section B of Supplementary Material. For notation simplicity, let $\delta := (\delta_{1},...,\delta_{T})$ with $\delta_{t} := (e_{t},\mu_{t},\tp_{t})$ denote the collection of all nuisance parameters, and let $\cT$ denote the space of $\delta$.

\begin{asu}[Convergence of nuisance parameter estimator]
$\hat{\delta}$ converges in $L_{2}$ to some limit $\delta' \in\cT$ as $n\to\infty$.
In other words, for each $t\in[T]$, there exists $\mu'_{t}(h_{t},a_{t})$ such that
$\|\hat{\mu}_{t}-\mu'_{t}\|=o_{p}(1)$; there exists $e'_{t}(h_{t},a_{t})$
such that $\|\hat{e}_{t}-e'_{t}\|=o_{p}(1)$; and there exists $\tp'_{t}(s_{t},a_{t})$
such that $\|\hat\tp_{t}-\tp'_{t}\|=o_{p}(1)$.\label{assu:(Convergence-of-nuisance) stated in the main paper}
\end{asu}
\begin{asu}[Unique zero]
There exists unique $\beta\in\Theta$ such that $\EE\{\tilde{U}(\beta,\delta')\}=0$.
\label{assu: unique zero stated in the main paper}
\end{asu}
\begin{asu}[Correctly specifying one nuisance parameter]
For each $t \in [T]$, suppose either $e'_{t}=e_{t}^{\star}$ or
$\mu'_{t}=\mu_{t}^{\star}$.\label{assu:(Double-robustness) stated in the main paper}
\end{asu}
\begin{asu}[Regularity conditions]
\label{assu:(Regularity-conditions) stated in the main paper} 
~
    \begin{asulist}
    \item \label{assu: regularity condition (i) parameter space is compact stated in the main paper} Suppose the parameter space $\Theta$ of $\beta$ is compact.
    \item \label{assu: regularity condition (ii) Support is bounded stated in the main paper} Suppose the support of $O$ is bounded.
    \item \label{assu: regularity condition (iii) invertibility of derivative stated in the main paper} Suppose $\EE{\{\partial_{\beta}U(\beta^{\star},\delta')\}}$
is invertible.
    \item \label{assu: regularity condition (iv) Donsker Condition stated in the main paper} For each $t$, all possible $e_{t}(\cdot)$,
$\mu_{t}(\cdot)$, and $\tilde{p}_{t}(\cdot)$ consists of a Donsker class.
    \item \label{assu: ehat, ebar are bounded away from 0 and 1 stated in the main paper} For each $t$,
$e'_{t}$ and $\hat{e}_{t}\in[\epsilon,1-\epsilon]$ for $\epsilon>0$.
    \end{asulist}
\end{asu}

We now present two versions for the asymptotic theory of the proposed $\hat\beta$, where the nuisance parameters $\hat{e}_{t}$ and $\hat{\mu}_{t}$ in Stage 1 of Algorithm \ref{algo:est} are either estimated parametrically (Theorem \ref{thm: CAN of parametric estimation with continuous outcome in the main paper}) or nonparametrically (Theorem \ref{thm: CAN of nonparametric estimation with continuous outcome in the main paper}). Whether $\tp_t$ is estimated parametrically or nonparametrically does not affect the asymptotic result because it is not needed to identify $\beta$ \citep{cheng2023efficient, lok2024estimating}.

\begin{thm}[Asymptotic normality with parametrically fitted nuisance function]
\label{thm: CAN of parametric estimation with continuous outcome in the main paper}
Suppose $e = (e_1,\ldots,e_T)$ and $\mu = (\mu_1, \ldots, \mu_T)$ are parameterized by finite-dimensional parameters $\gamma_e$ and $\gamma_\mu$, and suppose the estimators $\hat\gamma_e$ and $\hat\gamma_\mu$ solve estimating equations $\PP_nU_e(\gamma_e) = 0$ and $\PP_nU_\mu(\gamma_\mu) = 0$, respectively. Suppose Assumptions \ref{assu: consistency, Positivity, Sequential ignorability.} - \ref{assu:(Regularity-conditions) stated in the main paper} hold. Let $\theta := (\beta^T, \gamma_e^T, \gamma_\mu^T)^T$ and $\Phi(\theta, \tp) := (\tilde{U}(\theta, \tp)^T, U_e(\gamma_e)^T, U_e(\gamma_e)^T)^T$. Then $\hat\beta$ from Algorithm 1 is consistent and asymptotically normal: $\sqrt{n}(\hat{\beta}-\beta^{\star})\xrightarrow{d}N(0,V_1)$
as $n\rightarrow\infty$. Furthermore, $V_1$ can be consistently estimated by the upper diagonal $p$ by $p$ block matrix of  
\begin{align*}
\PP_n\{\partial_{\theta}\Phi(\hat{\theta}, \hat\tp)\}^{-1}\PP_n\{\Phi(\hat{\theta}, \hat\tp)\Phi(\hat{\theta}, \hat\tp)^{T}\}\PP_n\{\partial_{\theta}\Phi(\hat{\theta}, \hat\tp)\}^{-1,T},
\end{align*}
where $\hat\theta = (\hat\beta^T, \hat\gamma_e^T, \hat\gamma_\mu^T)^T$.
\end{thm}

\begin{thm}[Asymptotic normality with nonparametrically fitted nuisance function]
\label{thm: CAN of nonparametric estimation with continuous outcome in the main paper}
Suppose either or both of $e$ and $\mu$ are estimated nonparametrically.
Suppose Assumptions \ref{assu: consistency, Positivity, Sequential ignorability.} -
\ref{assu:(Regularity-conditions) stated in the main paper} hold. In addition, assume that $\|\hat{e}-e^{{\star}}\|\ \|\hat{\mu}-\mu^{{\star}}\|=o_p(n^{-1/2})$.
Then $\hat\beta$ from Algorithm 1 is consistent and asymptotically normal: $\sqrt{n}(\hat{\beta}-\beta^{{\star}})\xrightarrow{d}N(0,V_2)$ as $n\rightarrow\infty$,
where
\[
V_2=\EE\{\partial_{\beta}\tilde{U}(\beta^{{\star}},\delta')\}^{-1}\EE\{\tilde{U}(\beta^{{\star}},\delta')\tilde{U}(\beta^{{\star}},\delta')^{T}\}\EE\{\partial_{\beta}\tilde{U}(\beta^{{\star}},\delta')\}^{-1,T},
\]
 and $V_2$ can be consistently estimated by 
\[
\PP_n\{\partial_{\beta}\tilde{U}(\hat{\beta},\hat{\delta})\}^{-1}\PP_n\{\tilde{U}(\hat{\beta},\hat{\delta})\tilde{U}(\hat{\beta},\hat{\delta})^{T}\}\PP_n\{\partial_{\beta}\tilde{U}(\hat{\beta},\hat{\delta})\}^{-1,T}.
\]
\end{thm}

\begin{rmk}
     The required condition $\| \hat{e}-e^{\star}\|\ \|\hat{\mu}-\mu^{{\star}}\|=o_p(n^{-1/2})$ was termed ``rate double-robustness'' in \citet{smucler2019unifying}, and similar conditions are standard in the causal inference literature where machine learning algorithms are used to fit nuisance parameters \citep[e.g.,][]{chernozhukov2018double}. Because the term $\| \hat{e}-e^{\star}\|\ \|\hat{\mu}-\mu^{\star}\|$ is a product, we have two chances at obtaining adequately fast convergence rate. A sufficient condition for $\| \hat{e}-e^{\star}\|\ \|\hat{\mu}-\mu^{{\star}}\|=o_p(n^{-1/2})$ is $\|\hat{e}-e^{\star}\| = o_p(n^{-1/4})$ and $\| \hat{\mu}-\mu^{\star}\| = o_p(n^{-1/4})$, which is obtainable by many data-adaptive prediction algorithms such as generalized additive models under regularity conditions and optimal smoothing, and highly adaptive lasso \citep{horowitz2009semiparametric, van2017generally, kennedy2016semiparametric}. For nonparametric estimators that may not satisfy the Donsker condition, Algorithm \ref{algo:est} can be revised to employ cross-fitting \citep{chernozhukov2018double} and avoid Assumption \ref{assu:(Regularity-conditions) stated in the main paper} (iv).
\end{rmk}

\section{Simulation}
\label{sec: Simulation}

Our simulations focus on the causal excursion effect with $\Delta = 1$. Here we present simulation results verifying Theorem \ref{thm: CAN of nonparametric estimation with continuous outcome in the main paper} with identity link function $g$ (continuous outcome). Additional simulation results for log link $g$ (binary outcome) and for verifying Theorem \ref{thm: CAN of parametric estimation with continuous outcome in the main paper} are in Sections C and D of Supplementary Material. 

We set the total number of decision points per individual to $T = 20$. For each individual, we sequentially generate $(Z_t,A_t,R_{t,1},R_{t,1}Y_{t,1})_{1 \leq t \leq T}$. The time-varying covariate is exogeneous (independent of past history) and is generated from $Z_t \sim \text{Unif}(-2, 2)$. The treatment $A_t$ is generated from a Bernoulli distribution with a constant success probability $p_t(H_t) = 0.4$. The missingness indicator $R_{t,1}$ is generated from a Bernoulli distribution with success probability $e_t^\star(H_t, A_t)$. The proximal outcome was generated by $Y_{t, 1} = \mu^\star_t(H_t, A_t) + \xi_t$, where $\xi_t \sim N(0,1)$. The mean outcome is $\mu^\star_t(H_t, A_t) = A_t(\beta_0 + \beta_1 Z_t) + \mu^\star_t(H_t, 0)$. Each of $\mu^\star_t(H_t, 0)$ and $\text{logit} \{e_t^\star(H_t, A_t)\}$ follows one of three patterns: linear, simple nonlinear, or periodic in $Z_t$ and $t$; the detailed functional forms are presented in Table \ref{tab: simulation functions}.

\begin{table}
\centering
\begin{tabular}{c c c c} 
         \toprule
         & $\text{logit} e_t^\star(H_t, A_t)$  & $\mu^\star_t(H_t, 0)$ &  $g(\alpha_0, \alpha_1)$ \\
        \midrule
        Linear   & $g(-0.5, 1.5)$ & $g(0.5, 1.5)$ &  $\alpha_0 + \alpha_1 (t/T + Z_t/6)$\\
    Simple nonlinear  & $g(-2, 1.5)$ & $g(0.5, 1.5)$ & $\alpha_0 + \alpha_1\{q_{2,2}(Z_t/6 + 1/2) + q_{2,2}(t/T)\}$\\
    Periodic   & $g(0.5, 1.5)$ & $g(0.5, 1.5)$ & $\alpha_0 + \alpha_1\{\sin(t) + \sin(Z_t) \}$\\
         \bottomrule
  \end{tabular}
  \caption{Three data generating functions for $\mu^\star_t(H_t, 0)$ and $\text{logit} \{e_t^\star(H_t, A_t)\}$ in simulations. The generating functions are defined through a $g(\cdot,\cdot)$ function, which is defined in the last column. $q_{2,2}(\cdot)$ denotes the density function of Beta distribution.}
      \label{tab: simulation functions}
\end{table}

In the simulation, the goal is to estimate the causal excursion effect conditional on $Z_t$, $\cee_{\bp;\Delta = 1}(Z_t) = \beta_0 + \beta_1 Z_t$, with true parameter values set to $\beta_0 = 1.5$ and $\beta_1 = 2.1$. For each simulation setup, we used four sample sizes: 50, 100, 150, 200, each with 1000 replications.

We consider four implementations of Algorithm \ref{algo:est} that differ by their ways of fitting $\mu_t$ and $e_t$, as listed in Table \ref{tab: model specification}. Each implementation specifies two generalized additive models, one for $\hat{e}_t(H_t,A_t)$ and the other for $\hat\mu_t(H_t,A_t)$, fitted using the \texttt{gam} function in R package \texttt{mgcv} \citep{wood2017generalized}. Implementation A correctly specifies both the missingness and the outcome regression models. Implementation B misspecifies the missingness model by leaving out the spline term for $Z_t$, $s(Z_t)$. Implementation C misspecifies the outcome regression model by leaving out $s(Z_t)$ and $A_t s(Z_t)$. Implementation D misspecifies both models.

\begin{table}
\centering
\begin{tabular}{c c c} 
         \toprule
        Implementation& \begin{tabular}{c} Model for $\hat{e}_t(H_t,A_t)$ \\ (missingness) \end{tabular} & \begin{tabular}{c} Model for $\hat\mu_t(H_t,A_t)$ \\ (outcome regression) \end{tabular} \\
        \midrule
    A & $R_{t,1} \sim s(Z_t) + s(t)$ & $Y_{t, 1} \sim A_t\{s(Z_t) + s(t)\} + s(Z_t) + s(t)$  \\
        B & $R_{t,1} \sim s(t)$ & $Y_{t, 1} \sim A_t\{s(Z_t) + s(t)\} + s(Z_t) + s(t)$\\
        C & $R_{t,1} \sim s(Z_t) + s(t)$ & $Y_{t, 1} \sim A_ts(t) + s(t)$ \\
        D & $R_{t,1} \sim s(t)$ & $Y_{t, 1} \sim s(Z_t) + s(t)$  \\
             \bottomrule
  \end{tabular}
  \caption{Four implementations of Algorithm \ref{algo:est} used in the simulation. Each implementation uses the generalized additive model (\texttt{gam} function) in R package \texttt{mgcv}. $s(\cdot)$ denotes penalized splines. The model for $\hat{e}_t(H_t,A_t)$ is fitted using option \texttt{family = binomial}.}
  \label{tab: model specification}
\end{table}

\begin{figure}
    \centering
    \includegraphics[width = 0.85\textwidth]{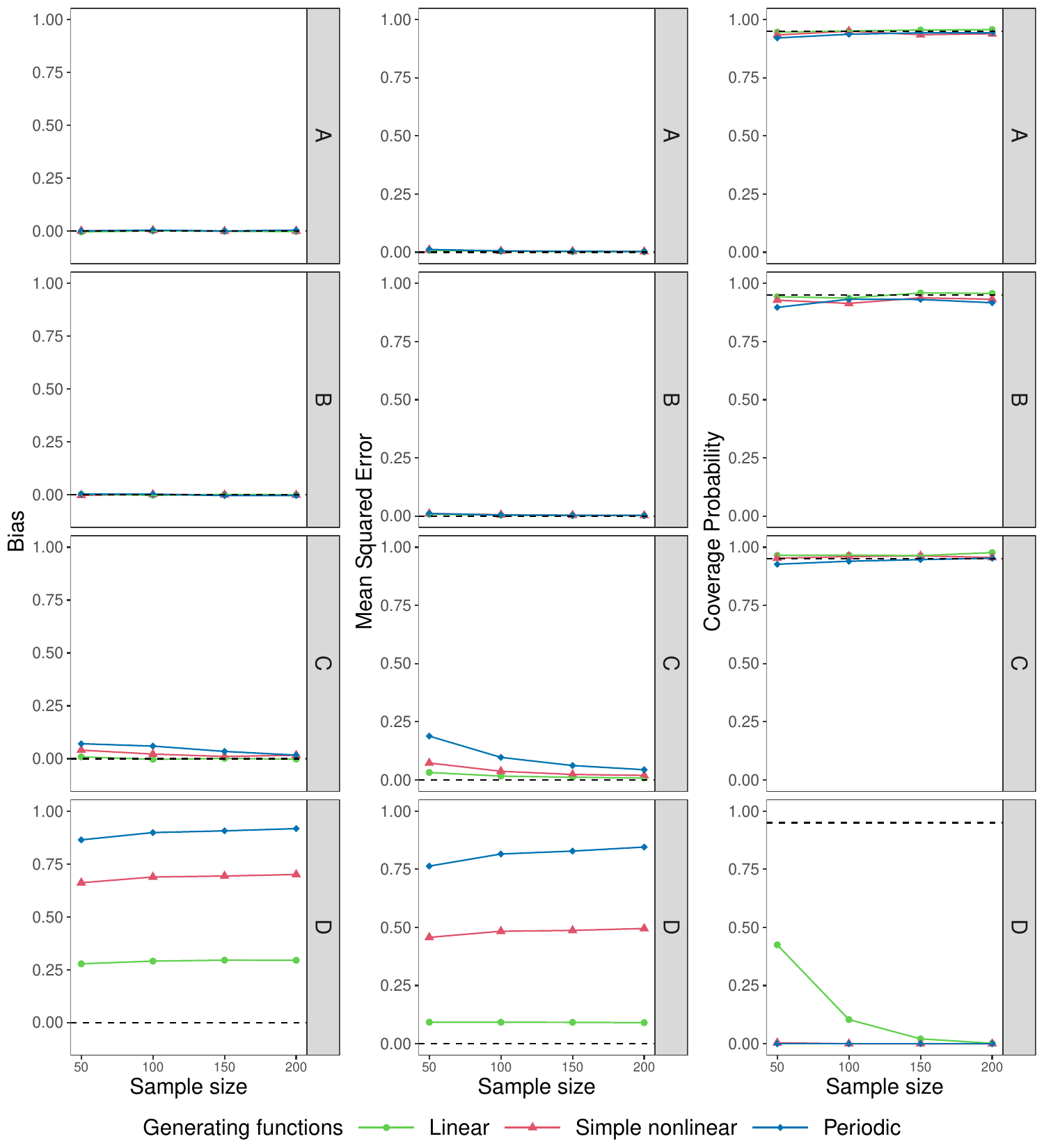}
    \caption{Bias, mean squared error, and coverage probability of $\hat \beta_0$ in simulation.}
    \label{fig: simulation results with continuous outcome with GAM for beta1}
\end{figure}

\begin{figure}
    \centering
    \includegraphics[width = 0.85\textwidth]{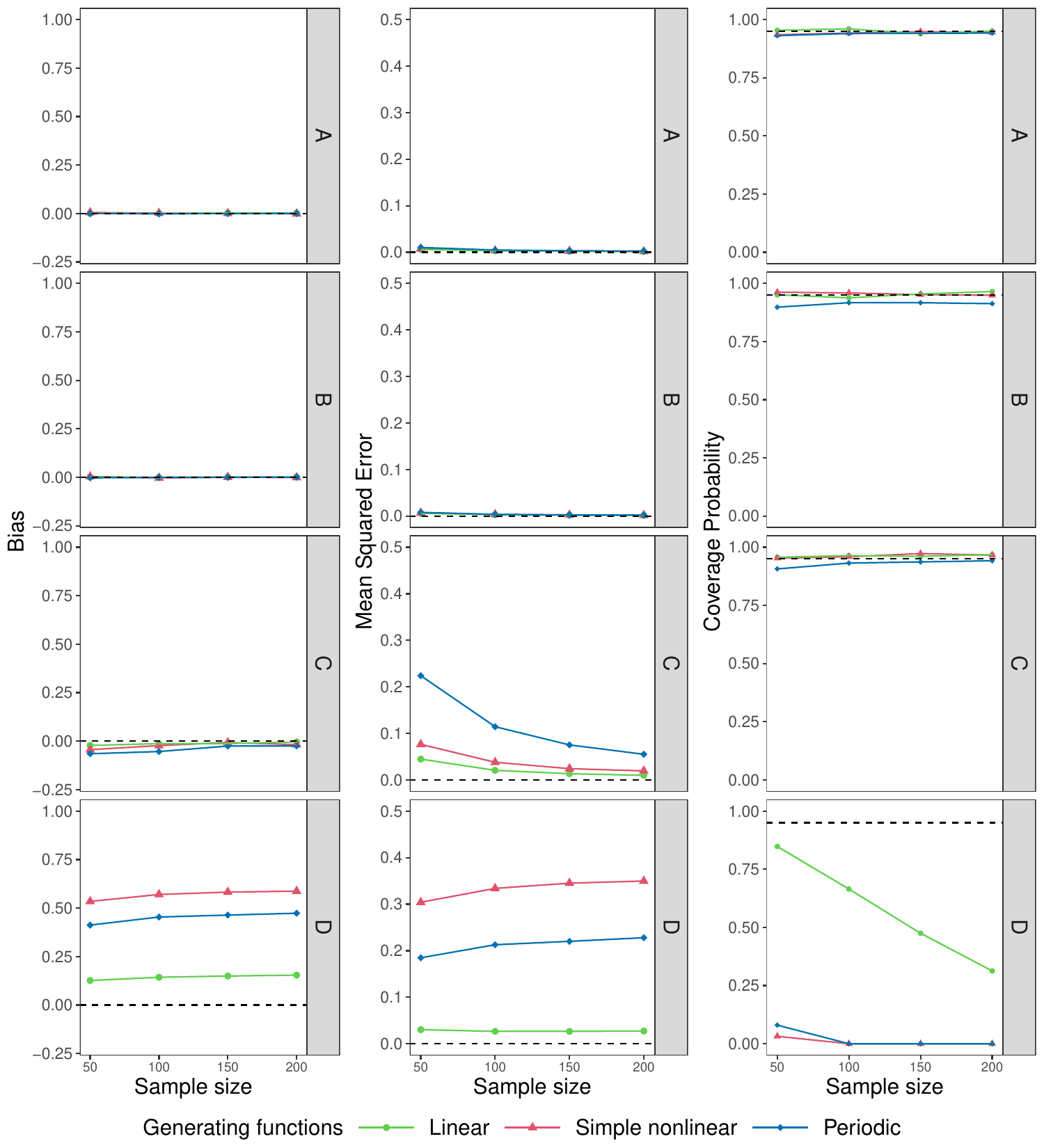}
    \caption{Bias, mean squared error, and coverage probability of $\hat \beta_1$ in simulation.}
    \label{fig: simulation results with continuous outcome with GAM for beta2}
\end{figure}

Figure \ref{fig: simulation results with continuous outcome with GAM for beta1} and \ref{fig: simulation results with continuous outcome with GAM for beta2} shows the bias, mean squared error (MSE) and coverage probability for $\hat \beta_0$ and $\hat \beta_1$ based on the four implementations. As the sample size increases, bias and MSE of $\hat \beta_0$ and $\hat \beta_1$ decreases for implementations A, B, and C. In addition, the coverage probability of 95\% confidence interval is close to the nominal level. This verifies the consistency and asymptotic normality result in Theorem \ref{thm: CAN of nonparametric estimation with continuous outcome in the main paper}. For implementation D, because neither the missingness model nor the outcome regression model is correctly specified, the estimator is inconsistent.

\section{Application}
\label{sec: Application}

We use data from HeartSteps I (hereafter HeartSteps; data available at \url{https://github.com/klasnja/HeartStepsV1/tree/main/data_files}), the first study in a series of HeartSteps MRTs aimed at designing mHealth interventions to promote physical activity among sedentary adults \citep{klasnja2019efficacy}. The study included 37 participants, each randomized up to five times per day over a 42-day period, totaling 210 decision points per participant. At each decision point, participants had a 60\% chance of receiving a smartphone-delivered physical activity suggestion. A participant was considered unavailable at a decision point if any of the following conditions were met: they were driving, had manually disabled the intervention, were exercising within 90 seconds of the decision point, or lacked an active internet connection \citep{seewald2019practical}.  Out of 7884 total decision points, 6331 (80.3\%) were available. Participants selected their preferred randomization times before the start of the study, ensuring at least 90-minute intervals between notifications. Each participant wore a
wristband tracker which continuously recorded their step count. The outcome of interest is the total step count in a subsequent 30-minute window following each decision point, and 9.4\% outcomes were missing.

To illustrate the method, we assessed the effect of activity suggestions on the 30-minute total step count. Specifically, we considered three estimands: (a) a fully marginal effect of the suggestion averaged over all decision points; (b) a moderated effect with the effect modifier being the decision point index; (c) a moderated effect with the effect modifier being whether the decision point is on a weekday. 

The Stage 1 nuisance functions $e_t(H_t, A_t)$ and $\mu_t(H_t, A_t)$ were fitted using either a generalized linear model or a generalized additive model with penalized splines. We used the following predictors to fit both $e_t(H_t, A_t)$ and $\mu_t(H_t, A_t)$, pooling across $t \in [T]$: the decision point index $t$, lag-1 outcome (the log-transformed step count in the 30-minute window preceding decision point $t$), location (an indicator for being at home or work), the weekday indicator, recent activity level (an indicator of whether the individual took fewer than 56 steps in the 30-minute window preceding decision point $t$), and interactions between treatment and each of these predictors. The numerator probability $\tp(S_t)$ is set to 0.6, the true randomization probability. The Stage 2 model was specified depending on the estimand. Wald-type 95\% confidence interval was obtained based on the asymptotic variance estimators in Theorems \ref{thm: CAN of parametric estimation with continuous outcome in the main paper} and \ref{thm: CAN of nonparametric estimation with continuous outcome in the main paper}.

We considered three comparator methods: weighted and centered least squares \citep[WCLS;][]{boruvka2018assessing} applied to only complete cases, WCLS after imputing missing outcomes as zero, and WCLS after imputing missing outcomes using individual-specific mean. We used the \texttt{wcls} function implemented in the \texttt{MRTAnalysis} R package \citep{tianchenpackage}. Recall that WCLS is appropriate for estimating the CEE had there been no missing outcomes. 

Table \ref{tab: DA continuous outcome result table in the main paper} shows the analysis result. Delivering an activity suggestion versus no suggestion increased the step count significantly, but the effect deteriorated over time. The results were qualitatively similar across the five methods (two proposed and three comparators) despite quantitative differences. This is likely due to the relatively low proportion of missing data in this dataset (9.4\%). However, the theoretical advantages of our approach such as the double robustness and the weaker assumption on missing data than its comparators make it appropriate in more settings. In datasets with higher proportions of missing outcomes, our method is expected to offer greater improvements in precision and validity. 

We also illustrate our method for a binary outcome of whether a participant takes more than 93 steps (population median) or not in the 30-minute window following each decision point. The observations are qualitatively similar to the continuous outcome illustration. Details can be found in the Section E of Supplementary Material.

\begin{table}[htbp]
\scriptsize
\centering
\begin{tabular}{c c c c c}
    \toprule
         Estimand &  Method & Intercept & Slope (decision point index) & Slope (is.weekday)\\
         \midrule
         \multirow{6}{*}{(a)} & DR-parametric & $0.14 \ (-0.01, 0.29)$ &  &  \\
         & DR-nonparametric & $0.15 \ (0.00, 0.29)$ &  &  \\
         & Complete-case & $0.15 \ (0.00, 0.30)^\star$ &  &  \\
         & Impute-zero & $0.13 \ (0.00, 0.27)$ &  &  \\ 
         & Impute-mean & $0.13 \ (0.00, 0.27)^\star$ &  &  \\
         \midrule
        \multirow{6}{*}{(b)} & DR-parametric & $0.47 \ (0.18, 0.76)^\star$ & $-0.003 \ (-0.005, -0.001)^\star$ &  \\
        & DR-nonparametric & $0.47 \ (0.18, 0.77)^\star$ & $-0.003 \ (-0.005, -0.001)^\star$ &  \\
         & Complete-case & $0.45 \ (0.15, 0.75)^\star$ & $-0.003 \ (-0.005, -0.001)^\star$ &  \\
         & Impute-zero & $0.45 \ (0.16, 0.73)^\star$ & $-0.003 \ (-0.005, -0.001)^\star$ &  \\ 
         & Impute-mean & $0.42 \ (0.12, 0.71)^\star$ & $-0.003 \ (-0.005, 0.000)^\star$  &  \\
         \midrule
         \multirow{6}{*}{(c)} & DR-parametric & $-0.10 \ (-0.36, 0.17)$ & & $0.33 \ (0.01, 0.65)^\star$  \\
        & DR-nonparametric & $-0.09 \ (-0.36, 0.18)$ & & $0.33 \ (0.01, 0.65)^\star$  \\
         & Complete-case & $-0.09 \ (-0.35, 0.17)$ & & $0.33 \ (0.02, 0.64)^\star$  \\
         & Impute-zero & $-0.05 \ (-0.30, 0.20)$ & & $0.26 \ (-0.04, 0.55)$  \\ 
         & Impute-mean & $-0.11 \ (-0.36, 0.14)$ & & $0.34 \ (0.04, 0.64)^\star$  \\
    \bottomrule
    \end{tabular}
  \caption{Coefficient estimates and 95\% confidence intervals for the three estimands stated in Section \ref{sec: Application}: (a) a fully marginal effect of the suggestion averaged over all decision points; (b) a moderated effect with the effect modifier being the decision point index; (c) a moderated effect with the effect modifier being whether the decision point is on a weekday. ``DR-parametric'' represents our doubly robust estimator with generalized linear models in Stage 1 nuisance estimation; ``DR-nonparametric'' represents our doubly robust estimator with generalized additive models in Stage 1 nuisance estimation; ``Complete-case'' represents the WCLS applied to only complete case; ``Impute-zero'' represents WCLS after imputing missing outcomes as zero; ``Impute-mean'' represents WCLS after imputing missing outcomes using individual-specific mean. Asterisk denotes exclusion of zero from the confidence interval.}
      \label{tab: DA continuous outcome result table in the main paper}
\end{table}

\section{Discussion}
\label{sec: Discussion}

 We proposed a two-stage doubly robust estimator for the causal excursion effect in MRTs where the longitudinal outcomes are missing at random. In the first stage, the two nuisance parameters (missingness model and outcome regression) can be fitted either parametrically or nonparametrically. When fitted parametrically, only one of the two nuisance parameter models needs to be correctly specified. When fitted nonparametrically, methods with a slower convergence rate (e.g., $o_p(n^{-1/4})$) can be accommodated, which enables a wider range of nonparametric and machine learning nuisance estimators. We established the asymptotic theory and verified the finite sample performance of the estimator through simulations.

One direction for future work is to extend the current methodology to adopt ideas from recent advances in doubly robust estimation, which may further improve the estimator's performance when both nuisance models are misspecified or when the nuisance parameter estimator converges at a very slow rate. The higher-order influence function idea \citep{diaz2016second} and restricted empirical maximum likelihood estimation \citep{tan2010bounded} are promising future directions. Another interesting extension is to incorporate enhanced propensity score models \citep{cao2009improving}, which balances weights for each individual by adding a scalar parameter in the logistic regression. Lastly, one may incorporate the bias-reduced doubly robust estimation \citep{vermeulen2015bias} to reduce the estimator's sensitivity to model misspecification when both nuisance models are only mildly misspecified.

Another direction for future work is developing methods to handle missing data without relying on the assumption of missing at random (MAR). A key challenge with missing not at random (MNAR) data is the issue of identification---i.e., the parameter of interest may not be uniquely identified from the observed data \citep{yang2019causal}. Shadow variables can be a potential solution \citep{miao2024identification}.

A third direction is to allow for missing covariate or treatment data. This can occur in MRTs due to technical glitches such as app malfunctions or sensor failures. 

\section*{Acknowledgements}
The authors acknowledge the support of UCI ICS Research Award.

\section*{Supplementary Material}

Supplementary material includes theoretical proofs, including identifiability result and proofs of Theorem \ref{thm: CAN of parametric estimation with continuous outcome in the main paper} and \ref{thm: CAN of nonparametric estimation with continuous outcome in the main paper}. We also present additional simulation results, and the data analysis result for binary outcome.  The code for replicating the simulations and data analysis can be accessed at \url{https://github.com/jiaxin4/missingCEE}.

\bibliographystyle{apalike}
\bibliography{ref}


\end{document}